\documentclass[11pt]{asaproc}

\usepackage{amsmath,amsfonts,graphicx,natbib,times,rotating}

\newcommand{\diff}{{\sf d}}
\newcommand{\trace}{\mathop{\rm tr}}
\newcommand{\Expect}{\mathbb{E}}
\newcommand{\Var}{\mathbb{V}}
\newcommand{\bfx}{\mathbf{x}}

\begin{document}

\title{Power analysis for a linear regression model \\ when regressors are matrix sampled}
\author{Stanislav (Stas) Kolenikov$^*$ \and Heather Hammer%
\thanks{Abt Associates, 8405 Colesville Rd, Suite 300, Silver Spring, MD 20910. Corresponding author: \texttt{stas\_kolenikov@abtassoc.com}}}
\maketitle

\begin{abstract}%
Multiple matrix sampling is a survey methodology technique that randomly chooses a relatively small subset of items
to be presented to survey respondents for the purpose of reducing respondent burden. The data produced are missing
completely at random (MCAR), and special missing data techniques should be used in 
linear regression and other multivariate statistical analysis. 
We derive asymptotic variances of regression parameter estimates that allow us to conduct
power analysis for linear regression models fit to the data obtained via a multiple matrix sampling design.
The ideas are demonstrated with a variation of the Big Five Inventory of psychological traits.
An exploration of the regression parameter space
demonstrates instability of the sample size requirements, and substantial losses of precision with matrix-sampled
regressors. A simulation with non-normal data demonstrates the advantages of a semi-parametric multiple imputation
scheme.

\end{abstract}

\section{Introduction and motivation}

This work was conceived and carried out in the context of a task to reduce 
respondent burden in a mental health study. We were interested in a range of outcome
variables, and our analytical goal was fitting regression models to explain the mental health outcomes.
The instrument collects demographic explanatory variables,
as well as scores from the Big Five Inventory \citep{john:sriv:1999:big5},
% TODO: REF NEEDED ON BIG FIVE
a commonly used set of five psychological traits that are often found to be correlated
with behaviors and outcomes. We expected that multiple matrix sampling would allow
us to reduce the instrument length from over an hour to about 20--25 minutes.
A key component of sampling design, sample size determination,
will be based on a linear regression power analysis.
However, complexities of regression analysis with missing data required custom
derivations of power analyses, which is what this technical paper addresses.

\section{Regression setting}

Consider a regression analysis problem where an outcome $y$ is predicted by a set of
explanatory variables $x_1, \ldots, x_p$:

\begin{equation}
    y = \beta_0 + \beta_1 x_1 + \ldots + \beta_p x_p + \varepsilon
    \label{eq:regress}
\end{equation}

In the simplest possible case of no missing data and homoskedastic normal errors $\Var[\epsilon_i]=\sigma^2 \, \forall i$,
the maximum likelihood estimates are the OLS estimates

\begin{gather}
    \hat\beta_{\rm OLS} = (X'X)^{-1} X'Y;
    \quad \Var[ \hat\beta_{\rm OLS} ] = \sigma^2 (X'X)^{-1};
    \notag \\
    v[ \hat\beta_{\rm OLS} ] = s^2 (X'X)^{-1};
    \quad s^2 = \frac1n \sum_{i=1}^n (y_i - x_i'\hat\beta)^2
    \label{eq:ols}
\end{gather}

Inference on regression coefficients is based on normality of coefficient estimates,
$\hat\beta \sim N(\beta,\sigma^2(X'X)^{-1})$.

An unbiased estimate of $s^2$ can be obtained by changing the denominator (degrees of freedom)
from $n$ to $n-p$. The OLS estimates hold desirable properties in more general settings,
e.g., by dropping the normality requirement. When regressors $X$ are stochastic, the OLS estimates
and their variance estimates given above only have an asymptotic justification, and require
independence of regressors $X$ and errors $\varepsilon$. When the basic assumptions are violated,
sandwich-type or resampling variance estimates need to be used.

\subsection{Power analysis and sample size determination in regression setting}
\label{subsec:reg:power}

Power analysis and sample size determination are statistical tasks of addressing,
quantifying and controling type I error. In a typical power analysis problem,
a null hypothesis, $H_0: \theta \in \Theta_0$, and an alternative, $H_1: \theta \in \Theta_1$,
are formulated; a test statistic $t(X)$ is selected, for which a critical region of level $\alpha$ is specified.
E.g., assuming that high values of the test statistic indicate disagreement between the data
and the null, as is typical with $\chi^2$ or $F$-statistic tests common in regression models,
the rejection region would have the form $T_\alpha = [c,+\infty)$ so that
$\rm{Prob} [t(X) > c] \le \alpha$ when the true value of the parameter $\theta \in \Theta_0$.
Finally, power analysis addresses the issue of Type II error, i.e., $\rm{Prob} [t(X)\le c]$ under
the alternative $\theta \in \Theta_1$. While the null hypothesis typically represents
a simple hypothesis $\theta=\theta_0$ or a subset of reasonably small dimension, the alternative
is necessarily complex. Hence researchers often formulate a measure of \textit{effect size}
$\delta$ and consider power analysis for parameter values under the alternative that are
at least $\delta$ away from the specific value $\theta_0$ or the subset $\Theta_0$ in an appropriate metric.

As the power to reject the null typically grows with the sample size $n$, the task
of sample size determination is to find the value $n_\beta$ that guarantees a given level
of power $1-\beta$. While the size of the test is often taken to be $\alpha=5\%$,
% TODO: see, however, calls to reduce the common value to $\alpha=0.5\%$
the traditional type II error rate is $\beta=20\%$ leading to $1-\beta=80\%$ power.

While testing the location parameter of two populations, for example, the natural hypotheses
might be $H_0: \mu_1 = \mu_2$ vs. the (two-sided) alternative $H_1: |\mu_1 - \mu_2| \ge \delta$,
with relatively straightforward testing based on the Student $t$ distribution, linear regression
models feature a variety of statistics that may be subject to testing and power analysis:
\begin{enumerate}
    \item Test of overall fit: $H_0: \mathbf{\beta}=0$.
    \begin{enumerate}
        \item A version of this hypothesis can be formulated as $H_0: R^2=0$. Depending
            on how easy or difficult it is to conduct inference on parameter estimates
            or regression sums of squares, one or the other may be preferred in applications.
    \end{enumerate}
    \item An increase of overall fit: $H_0: R^2 \le R_0^2$ vs. $H_1: R^2 \ge R_0^2 + \delta$.
    \item Specific regression coefficients: the coefficient of the $j$-th explanatory variable
        is zero, $H_0: \beta_j = 0$ vs. $H_1: | \beta_j | \ge \delta$.
    \item Linear hypothesis $H_0: R\beta = r_0$, which covers cases like:
    \begin{enumerate}
        \item Equality of two regression coefficient for the $j$-th and the $k$-th explanatory variable,
        $H_0: \beta_j - \beta_k = 0$, so $R = (0, \ldots, 1, \ldots, -1), r=0$
        \item No impact of a set of variables $j_1, j_2, \ldots$: $H_0: \beta_{j_1}=0, \beta_{j_2}=0, \ldots$,
        so that $R$ is a subset of rows of a unit matrix, and $r$ is a zero vector of conforming dimension.
    \end{enumerate}
    \item Tests on error variance $\sigma^2$, e.g. $H_0: \sigma_2 \le \sigma^2_0$ vs. $H_1: \sigma^2 \ge \sigma_0^2 + \delta$.
\end{enumerate}

\section{Multiple matrix sampling}

\textit{Multiple matrix sampling} of a survey questionnaire consists
of administering only a specific subset of items to a given respondent, out of all items
this respondent is potentially eligible to be asked. The name stems from representation of the data
with respondents as rows, and items as columns, so that matrix sampling concerns selecting
specific entries in the matrix to be administered, rather than the full row as is typically done.
The focus of the technique is on selecting items out of all the relevant ones that the respondent
could be asked, with the potential skip  patterns already taken into account.
Similar or equivalent techniques are also known as \textit{partitioned designs}
and \textit{questionnaire splitting}. The method originated in educational testing
\citep{shoemaker:1972}, where it was first used to select items from a large pool
of available ones. The educational testing companies have identified the need to implement
multiple matrix sampling methods to protect the integrity of their data products, so that
the students taking a standardized test are not able to get
trained on a small subset of items known to be administered on standardized tests, thus
biasing the estimates of achievement.

Given the relatively esoteric nature of the method, the existing publications
have addressed some specific niche problems in matrix sampling (and the current paper
is no exception).
\citet{gonzalez:eltinge:2007} provided a review of matrix sampling
and applications in Consumer Expenditure Quarterly Survey.
\citet{chipperfield:steel:2009} put the problem of matrix sampling into a cost
optimization framework, where proper subsets of $K$ items can be administered on
up to $2^K-1$ forms at specific cost per item rates. They demonstrated that
with two items (or groups of items),
the split questionnaire best linear unbiased estimator (BLUE; also related to the GLS
estimator) provides modest efficiency gains over a design in which all items
are administered at once, and over a two-phase design in which all items are administered
to a fraction of respondents, and one subset of items is administered to the remainder of respondents.
\citet{merkouris:2010} extended their work to provide simplified composite estimation
using the estimates based on the form-specific subsamples, where compositing is based
on the second-order probabilities of selection and the way they are utilized
in estimating the variance of the Horvitz-Thompson estimator.
\citet{eltinge:2013} discussed connections to and relations with multiple frame
and multiple systems estimation methods (e.g., integration of survey and administrative
data, where administrative data may fill some of the survey items when available).
We add to this literature by providing the asymptotic variance-covariance matrix
of the coefficient estimates under matrix sampling of regressors, assuming
that the outcome is always collected. We also discuss implications for power analysis
and sample size determination.

\subsection{A simple example}

Consider the following matrix sampling design, in which
the outcome $y$ is collected on every form, while the explanatory variables
differ between forms.

\begin{table}[!h]

\centering

\caption{Three questionnaire forms for data collection: Design 1. \label{tab:forms3:design1}}

\begin{tabular}{l|ccc|c}
    Form & $X_1$ & $X_2$ & $X_3$ & $n$ \\
    \hline
    1   & + & & & $n_1$ \\
    2   & & + & & $n_2$ \\
    3   & & & + & $n_3$ \\
\end{tabular}

\end{table}

With this design, summaries (means, totals) of all the variables
($x_1,x_2,x_3,y$) can be obtained, and the bivariate relations
between each of the regressors and the outcome $y$ can be analyzed.
However, estimation of a multiple regression model requires
estimability of all of the entries of the $(X'X)$ matrix,
which this specific matrix sampling design does not provide.

To conduct regression analysis, we need to observe the cross-entries
of the $X'X$ matrix, which necessitates the following matrix sampling design.

\begin{table}[!h]

\centering

\caption{Three questionnaire forms for data collection: Design 2. \label{tab:forms3:design2}}

\begin{tabular}{l|ccc|c}
    Form & $X_1$ & $X_2$ & $X_3$ & $n$ \\
    \hline
    1   & & + & + & $n_1$ \\
    2   & + & & + & $n_2$ \\
    3   & + & + & & $n_3$ \\
\end{tabular}

\end{table}

\subsection{Parameter estimation under matrix sampling}

Since the components of $X'X$ and/or $X'y$ necessary to obtain the OLS regression
estimates may not be jointly available, more complex estimation strategies may need to
be employed. We study two such strategies.

One possibility is to utilize structural equation modeling (SEM) with missing data,
in which the marginal regression model of interest is formulated by using the regressors
as exogenous variables, the dependent variable is introduced as the only endogenous variable
explained by the model
% Bollen's demonstration of regression as an SEM method
\citep{bollen:selv:1989},
and the existing SEM estimation methods are applied
% Vika's missing data papers?
\citep{yuan:bentler:2000,savalei:2010}.

Alternatively, since the data are missing by design, and can be treated as MCAR, multiple imputation
\citep[MI]{rubin:1996,vanbuuren:2012}
% TODO: multiple imputation refs
can be used to fill in the missing values, with Rubin's variance formulae used to combine
MI estimates and provide inference. Of the several existing flavors of multiple imputation, 
one of the simplest strategies is
imputation under multivariate normality (which we expect to behave in ways similar to
the estimation methods for SEM with missing data under multivariate normality).
A less model-dependent method is predictive mean matching \citep{little:1988}
% TODO: reference on pmm? van Buuren's book?
in which a regression model is fit for each imputed variable, a linear prediction
is obtained for each case with missing variable, and an imputation is made by choosing
the value of the dependent variable from one of the nearest neighbors in terms of
the linear prediction score.

\section{Set up and notation}

All of the derivations in this paper concern the joint matrix of the first and second order moments of the data:
\begin{equation}
    \Omega =
    \Expect
    \left[
    \begin{pmatrix}
      1 \\
      \bfx \\
      y
    \end{pmatrix}
    \begin{pmatrix}
      1 &
      \bfx' &
      y
    \end{pmatrix}
    \right]
    =
    \begin{pmatrix}
      1 & \mu_\mathbf{x}' & \mu_y \\
      \mu_\mathbf{x} & \Expect[ \bfx \bfx'] & \Expect[\bfx y] \\
      \mu_y & \Expect[\bfx' y] & \Expect[y^2]
    \end{pmatrix}
    \equiv
    \begin{pmatrix}
      \omega_{00} & \Omega_{0x} & \omega_{0y} \\
      \Omega_{0x}' & \Omega_{xx} & \Omega_{xy} \\
      \omega_{0y} & \Omega_{xy}' & \omega_{yy}
    \end{pmatrix}
    \label{eq:Omega}
\end{equation}

The maximum likelihood estimates of the coefficients in the regression of $y$ on $x$
(obtained, for instance, through SEM modeling using maximum likelihood estimates with multivariate normal missing data method;
or approximated through multiple imputation) are obtained as
\begin{equation}
    \hat\beta_{\rm FIML} =
    \begin{pmatrix}
        \omega_{00} & \hat\Omega_{0x} \\
        \hat\Omega_{0x}' & \hat\Omega_{xx}
    \end{pmatrix}^{-1}
    \begin{pmatrix}
        \hat\omega_{0y} \\
        \hat\Omega_{xy}
    \end{pmatrix}
    \label{eq:ml:regression}
\end{equation}
where $\hat\Omega$ is the maximum likelihood estimator of the joint parameter matrix:
\begin{equation}
    \hat\Omega =
    \begin{pmatrix}
        \omega_{00} & \hat\Omega_{0x} & \hat\omega_{0y} \\
        \hat\Omega_{0x}' & \hat\Omega_{xx} & \hat\Omega_{xy} \\
        \hat\omega_{0y} & \hat\Omega_{xy}' & \hat\omega_{yy}
    \end{pmatrix}
    =
    \begin{pmatrix}
            \omega_{00} & \hat\omega_{01} & \ldots & \hat\omega_{0p} & \hat\omega_{0y}  \\
        \hat\omega_{01} & \hat\omega_{11} & \ldots & \hat\omega_{1p} & \hat\omega_{y1}  \\
        \vdots & \ldots & \vdots & \vdots & \vdots  \\
        \hat\omega_{0p} & \hat\omega_{1p} & \ldots & \hat\omega_{pp} & \hat\omega_{yp}  \\
        \hat\omega_{0y} & \hat\omega_{y1} & \ldots & \hat\omega_{yp} & \hat\omega_{yy}
    \end{pmatrix}
    \label{eq:hat:Sigma}
\end{equation}
where $x_0=1$ is the regression intercept by convention, so that $\omega_{00}\equiv 1$, $\hat\omega_{0j}=\hat\mu_j$
are the (estimated) means of the $j$-th explanatory variable, and $\hat\omega_{0y}=\hat\mu_y$ is the estimated mean of $y$.

To derive the likelihood, we need the form-specific submatrices obtained by multiplying the
overall matrix by selector matrices. For instance, in Design 2 above, for the first form, the relevant covariance matrix is
\begin{equation}
    {\rm Cov}(x_2, x_3, y)' = F_1 \Omega F_1',
    \quad
    F_1 =
    \begin{pmatrix}
        0 & 0 & 1 & 0 & 0 \\
        0 & 0 & 0 & 1 & 0 \\
        0 & 0 & 0 & 0 & 1
    \end{pmatrix}
    \label{eq:selector:1}
\end{equation}
Matrices necessary to form $F_2\Omega F_2'$ and
$F_3 \Omega F_3'$ are defined in a similar way.

Define the unit selector vector
that picks up the estimates of the means $e_0=(1,0,\ldots,0)$, which is the unit vector with 1
in the ``zeroth'' position corresponding to the intercepts in the parameter matrix $\Omega$.
In addition to $e_0$ selecting the first order moments, define the unit selection vectors
$e_y=(0,0,0,0,1)'$ as the unit vector selecting the last row/column of $\Omega$ corresponding to the $y$-parameters,
and $e_j=(0,\ldots,0,1,0,\ldots)$ is a unit vector with 1 in the $j$-th position corresponding to the $j$-th variable
(with the convention of indexing starting at zero). Then we observe that
\begin{align}
    F_1' & = (e_2, e_3, e_y) \notag \\
    F_2' & = (e_1, e_3, e_y) \notag \\
    F_3' & = (e_1, e_2, e_y)
\end{align}

\section{Likelihood and derivatives}

\subsection{Likelihood}

Indexing the forms by $k$, and observations within forms by $i$, the likelihood can be written as
\begin{align}
    \ln L(\omega;X) =
        \sum_{k=1}^3 \sum_{i=1}^{n_k} &
            \Bigl\{
                - \frac12 \trace (F_k F_k') \ln(2\pi)
                - \frac12 \ln \det ( F_k \Omega F_k' )
                \notag \\ &
                - \frac12 \bigl[ (x_i',y) - e_0' \Omega F_k' \bigr] (F_k \Omega F_k')^{-1}
                        \bigl[ (x_i', y)' - F_k \Omega e_0 \bigr]
            \Bigr\}
    \label{eq:log:lkhd}
\end{align}
where $n_k$ is the number of observations on which the $k$-th form is collected,
and $F_k$ is the selector matrix corresponding to the $k$-th form.

Derivations of the asymptotic properties of the MLE estimate $\widehat\Omega$
are based on the matrix differential \citep{magn:neud:1999}
\begin{align}
    \diff \Omega = &
        \diff \omega_{00} \, e_0 e_0'
        + \sum_{j=1}^p \diff \omega_{0j} \, (e_0 e_j' + e_j e_0')
        + \diff \omega_{0y} \, (e_0 e_y' + e_y e_0')
        + \sum_{j=1}^p \diff \omega_{jj} \, e_j e_j'
    \notag \\ &
        + \sum_{j=1}^p \sum_{i\neq j} \diff \omega_{ij} \, (e_i e_j' + e_j e_i')
        + \sum_{j=1}^p \diff \omega_{yj} \, (e_y e_j' + e_j e_y'),
    \label{eq:omega:as:sum:of:e:crossed}
\end{align}

After some tedious algebra,
the following information matrix $\Expect \nabla^2 \ln L(\omega;X)$ results.

\begin{align}
    \Expect \Bigl[ \frac {\partial^2 \ln L(\omega;X)}{\partial \omega_{0s}\partial \omega_{0t}} \Bigr]
    & =
    - \sum_{k=1}^3 n_k \tau^{(k)}_{st},
    \label{eq:d2lnl:domega0:domega0}
    \\
    \Expect \Bigl[ \frac {\partial^2 \ln L(\omega;X)}{\partial \omega_{0s}\partial \omega_{uu}} \Bigr]
    & = 0,
    \label{eq:d2lnl:domega0:domega:uu}
    \\
    \Expect \Bigl[ \frac {\partial^2 \ln L(\omega;X)}{\partial \omega_{0s}\partial \omega_{uv}} \Bigr]
    & = 0
    \label{eq:d2lnl:domega0:domega:uv}
\end{align}

The zero expected cross-derivatives indicate that the estimates of the multivariate
normal means and the variance-covariance parameters are independent. (This may not be the case in general
if the missing data mechanism coded by the matrices $F_k$ is not MCAR, and instead related to the data values.)

\begin{align}
    \Expect \Bigl[ \frac {\partial^2 \ln L(\omega;X)}{\partial \omega_{ss}\partial \omega_{uu}} \Bigr]
        & = - \frac12 \sum_{k=1}^3 n_k \bigr[ \tau^{(k)}_{su} \bigl]^2
    \label{eq:d2lnl:domega:ss:domega:uu}
    \\
    \Expect \Bigl[ \frac {\partial^2 \ln L(\omega;X)}{\partial \omega_{ss}\partial \omega_{uv}} \Bigr]
    & = - \sum_{k=1}^3 n_k \tau^{(k)}_{su} \tau^{(k)}_{sv}
    \label{eq:d2lnl:domega:ss:domega:uv}
    \\
    \Expect \Bigl[ \frac {\partial^2 \ln L(\omega;X)}{\partial \omega_{st}\partial \omega_{uv}} \Bigr]
    & = - \sum_{k=1}^3 n_k \bigl[ \tau^{(k)}_{su} \tau^{(k)}_{tv} + \tau^{(k)}_{sv} \tau^{(k)}_{tu} \bigr]
    \label{eq:d2lnl:domega:st:domega:uv}
    \\
    \tau^{(k)}_{st} & = e_s' F_k' (F_k \Omega F_k')^{-1} F_k e_t
    \label{eq:tau:kst}
\end{align}
where
$\tau^{(k)}_{st}$ is the $(s,t)$-th entry of the inverse of the form-specific covariance matrix;
% $\zeta_{ks,i}$ is  a GLS residual for the variable indexed $s$ in the form /
% missing data pattern $k$, with observation index $i$ nested in $k$, scaled by its variance;
and indices $s,t,u,v$ can enumerate the explanatory variables $x_j$ and the response $y$.
As $x_i$ and $y$ are considered jointly multivariate normal at this point, there is no separation into
dependent and explanatory variables.

Putting these entries together into a matrix, and using
the standard maximum likelihood estimation theory results, the asymptotic variance
of the maximum likelihood estimates of $\mathop{\rm vech}\Omega$ is given by
\begin{equation}
    {\rm As.} \Var[ \hat\omega ] = - \Expect \bigr[ \nabla^2 \ln L(\omega;X) \bigr]^{-1}
    \label{eq:asvar:omega-hat}
\end{equation}

\subsection{The delta method derivation of the asymptotic variance of $\hat\beta$}

Let us now return to the task of estimating the coefficients of the regression equation
$$y=\beta'x + \epsilon$$
via (\ref{eq:ml:regression}). The asymptotic variance-covariance matrix of $\hat\beta_{\rm FIML}$
can be obtained from the asymptotic covariance matrix of $\hat\Omega$ using the delta-method,
i.e., linearization of the relation (\ref{eq:ml:regression}):
\begin{align}
    \diff \beta = &
    -
    \begin{pmatrix}
        \omega_{00} & \Omega_{0x} \\
        \Omega_{0x}' & \Omega_{xx}
    \end{pmatrix}^{-1}
    \begin{pmatrix}
        0 & \diff \Omega_{0x} \\
        \diff \Omega_{0x}' & \diff \Omega_{xx}
    \end{pmatrix}
    \begin{pmatrix}
        \omega_{00} & \Omega_{0x} \\
        \Omega_{0x}' & \Omega_{xx}
    \end{pmatrix}^{-1}
    \begin{pmatrix}
        \omega_{0y} \\
        \Omega_{xy}
    \end{pmatrix}
    +
    \notag \\
    & \hspace{5cm} +
    \begin{pmatrix}
        \omega_{00} & \Omega_{0x} \\
        \Omega_{0x}' & \Omega_{xx}
    \end{pmatrix}^{-1}
    \begin{pmatrix}
        \diff \omega_{0y} \\
        \diff \Omega_{xy}
    \end{pmatrix}
    \label{eq:diff:beta}
\end{align}
where the individual components of $\diff\Omega$ can be obtained from
(\ref{eq:omega:as:sum:of:e:crossed}).
Thus
\begin{align}
    \frac{\partial\beta}{\partial\omega_{0j}} = &
        -
        \begin{pmatrix}
            \omega_{00} & \Omega_{0x} \\
            \Omega_{0x}' & \Omega_{xx}
        \end{pmatrix}^{-1}
        \begin{pmatrix}
            0 & e_j' \\
            e_j & \underline{0}
        \end{pmatrix}
        \begin{pmatrix}
            \omega_{00} & \Omega_{0x} \\
            \Omega_{0x}' & \Omega_{xx}
        \end{pmatrix}^{-1}
        \begin{pmatrix}
            \omega_{0y} \\
            \Omega_{xy}
        \end{pmatrix}
    \notag \\
    \frac{\partial\beta}{\partial\omega_{0y}} = &
    \begin{pmatrix}
        \omega_{00} & \Omega_{0x} \\
        \Omega_{0x}' & \Omega_{xx}
    \end{pmatrix}^{-1}
    \begin{pmatrix}
        1 \\
        \vec{0}
    \end{pmatrix}
    \notag \\
    \frac{\partial\beta}{\partial\omega_{jj}} = &
    -
    \begin{pmatrix}
        \omega_{00} & \Omega_{0x} \\
        \Omega_{0x}' & \Omega_{xx}
    \end{pmatrix}^{-1}
    \begin{pmatrix}
        0 & \vec{0}' \\
        \vec{0} & e_j e_j'
    \end{pmatrix}
    \begin{pmatrix}
        \omega_{00} & \Omega_{0x} \\
        \Omega_{0x}' & \Omega_{xx}
    \end{pmatrix}^{-1}
    \begin{pmatrix}
        \omega_{0y} \\
        \Omega_{xy}
    \end{pmatrix}
    \notag \\
    \frac{\partial\beta}{\partial\omega_{ij}} = &
    -
    \begin{pmatrix}
        \omega_{00} & \Omega_{0x} \\
        \Omega_{0x}' & \Omega_{xx}
    \end{pmatrix}^{-1}
    \begin{pmatrix}
        0 & \vec{0}' \\
        \vec{0} & e_i e_j' + e_j e_i'
    \end{pmatrix}
    \begin{pmatrix}
        \omega_{00} & \Omega_{0x} \\
        \Omega_{0x}' & \Omega_{xx}
    \end{pmatrix}^{-1}
    \begin{pmatrix}
        \omega_{0y} \\
        \Omega_{xy}
    \end{pmatrix}
    \notag \\
    \frac{\partial\beta}{\partial\omega_{yj}} = &
    \begin{pmatrix}
        \omega_{00} & \Omega_{0x} \\
        \Omega_{0x}' & \Omega_{xx}
    \end{pmatrix}^{-1}
    \begin{pmatrix}
        0 \\
        e_j
    \end{pmatrix}
    \notag \\
    \nabla_\omega \beta = & \Bigl(
            \frac{\partial\beta}{\partial\omega_{01}}, \frac{\partial\beta}{\partial\omega_{02}}, \ldots,
            \frac{\partial\beta}{\partial\omega_{y0}},
            \frac{\partial\beta}{\partial\omega_{11}}, \frac{\partial\beta}{\partial\omega_{12}}, \ldots, \frac{\partial\beta}{\partial\omega_{y1}}, \frac{\partial\beta}{\partial\omega_{22}}, \ldots,
            \frac{\partial\beta}{\partial\omega_{yp}}, 0 \Bigr)
    \label{eq:dbeta:domega}
\end{align}
where the derivatives are with respect to the components of the vectorization $\mathop{\rm vech}\Omega$,
of which the last term is $\frac{\partial\beta}{\partial\omega_{yy}}=0$. By the standard multivariate
delta-method results \citep{newey:mcfadden:1994,vandervaart:1998},
\begin{equation}
    {\rm As.} \Var[ \hat\beta ] = \nabla_\omega \beta \, \Var[ \hat\omega ] \, \nabla_\omega' \beta
    \label{eq:asvar:beta-hat}
\end{equation}

\section{Example: Big Five Inventory}

In our application, we wanted to analyze the relation between mental health outcomes
and the Big Five personal traits:

\begin{itemize}
    \item Openness to experience (inventive/curious vs. consistent/cautious)
    \item Conscientiousness (efficient/organized vs. easy-going/careless)
    \item Extraversion (outgoing/energetic vs. solitary/reserved)
    \item Agreeableness (friendly/compassionate vs. challenging/detached)
    \item Neuroticism (sensitive/nervous vs. secure/confident)
\end{itemize}

These personal traits have been found in numerous studies to be related to academic performance, disorders, general health,
and many other behaviors and outcomes. The standard Big Five scale consists of 44 items,
some of which are reverse worded and reverse scored to minimize the risk of straightlining, and with
items from different subscales mixed throughout the scales. Each item is a 5 point Likert scale with a clear midpoint.

In the population of interest, the Big Five traits are expected to have the following correlations, based
on preceding research:

\begin{equation}\label{eq:bigfive:corr}
  {\rm Cov}[x] =
  \begin{pmatrix}
    1    & 0.26 & 0.47 & 0.20 & -0.16 \\
    0.26 & 1    & 0.28 & 0.46 & -0.28 \\
    0.47 & 0.28 & 1    & 0.20 & -0.35 \\
    0.20 & 0.46 & 0.20 & 1    & -0.37 \\
    -0.16 & -0.28 & -0.35 & -0.37 & 1
  \end{pmatrix}
  \equiv
  \Sigma_{\rm Big 5}
\end{equation}

We thus consider a regression model
$$
    y_i = \beta_0 + \beta_1 x_{i1} + \ldots \beta_5 x_{i5} + \varepsilon_i
$$
where $x_{i1},\ldots,x_{i5}$ are subscale scores of the Big Five traits.
Measurement error in these scores is ignored, although more accurate methods are
available to account for it \citep{skrondal:laake:2001}.

A balanced multiple matrix sampling design would consist of ten forms,
each administering the outcome $y$ and two of the Big Five subscales:

\begin{table}[!h]

\centering

\caption{Multiple matrix sampling design with five explanatory variables. \label{tab:bigfive}}

\medskip

\begin{tabular}{l|cccccccccc}
    Form & 1 & 2 & 3 & 4 & 5 & 6 & 7 & 8 & 9 & 10 \\
    \hline
       O $\vphantom{\Bigl|}$ & + & + & + & + &   &   &   &   &   &    \\
       C $\vphantom{\Bigl|}$ & + &   &   &   & + & + & + &   &   &    \\
       E $\vphantom{\Bigl|}$ &   & + &   &   & + &   &   & + & + &    \\
       A $\vphantom{\Bigl|}$ &   &   & + &   &   & + &   & + &   & +  \\
       N $\vphantom{\Bigl|}$ &   &   &   & + &   &   & + &   & + & +  \\
        $y\vphantom{\Bigl|}$ & + & + & + & + & + & + & + & + & + & +  \\
\end{tabular}

\end{table}

\section{Simulation 1: Parameter space exploration}
\label{sec:explore}

In this simulation exercise, we explore the parameter space of
regression coefficients to gauge the degree of variability of sample size
determination results. Asymptotic variance resulting from
(\ref{eq:asvar:beta-hat}) is used to obtain the sample sizes
for the tasks outlined in Section \ref{subsec:reg:power}.
Simulation 1 consists of the following steps.

\begin{enumerate}
    \item Population regression parameters are simulated from $\beta \sim N(\mathbf{0},I_5)$.
    \item To provide the scale of the residual variance, the fraction of explained variance is set to
        $R^2=0.15$, a moderate effect for behavioral and social science data, and the associated
        residual variance $\sigma_\varepsilon^2$ is calculated based on this value of $R^2$.
    \item The complete data variances stemming from (\ref{eq:ols}) are recorded.
        \label{enum:explor:full:var}
    \item The multiple-matrix-sampled data variances stemming from (\ref{eq:asvar:beta-hat}) are recorded.
        \label{enum:explor:mcar:var}
    \item Sample size to reject the test of overall significance $H_0: \beta_1 = \ldots = \beta_5 = 0$
        at 5\% level with 80\% power is recorded.
        \label{enum:explor:n:overall}
    \item Sample size to detect an increase in $R^2$ by 0.01 (i.e., from 0.15 to 0.16), through a uniform
        multiplicative increase in the values of the regression parameters, keeping the residual variance
        $\sigma_\varepsilon^2$ constant, at 5\% level with 80\% power, is recorded.
        \label{enum:explor:n:infl}
    \item Sample size to detect an increase in $R^2$ by 0.01 (i.e., from 0.15 to 0.16), through an increase
        in the value of the coefficient $\beta_j, j=1, \ldots, 5$, keeping the residual variance
        $\sigma_\varepsilon^2$ constant and other regression parameters constant, at 5\% level with 80\% power, is recorded.
        \label{enum:explor:n:poke}
    \item Fraction of missing information (FMI) is computed as one minus the ratio
        of the variance of regression parameter estimate with complete data (obtained in step \ref{enum:explor:full:var})
        to the variance of regression parameter estimate with missing data (obtained in step \ref{enum:explor:mcar:var})
        \label{enum:explor:fmi}
\end{enumerate}

\begin{figure}[!bh]

    \centering

    \begin{tabular}{cc}
        \includegraphics[scale=0.09]{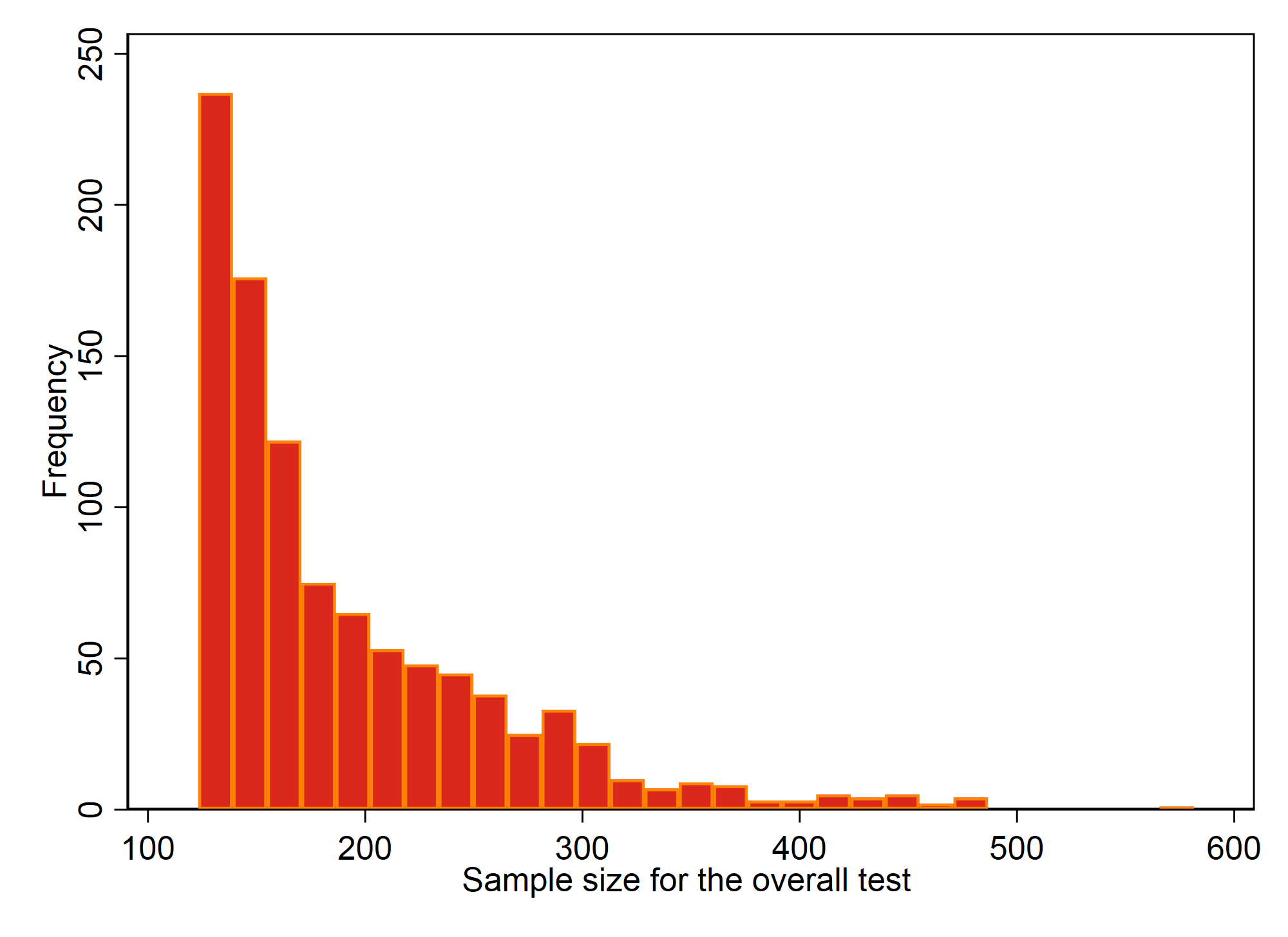}
        &
        \includegraphics[scale=0.09]{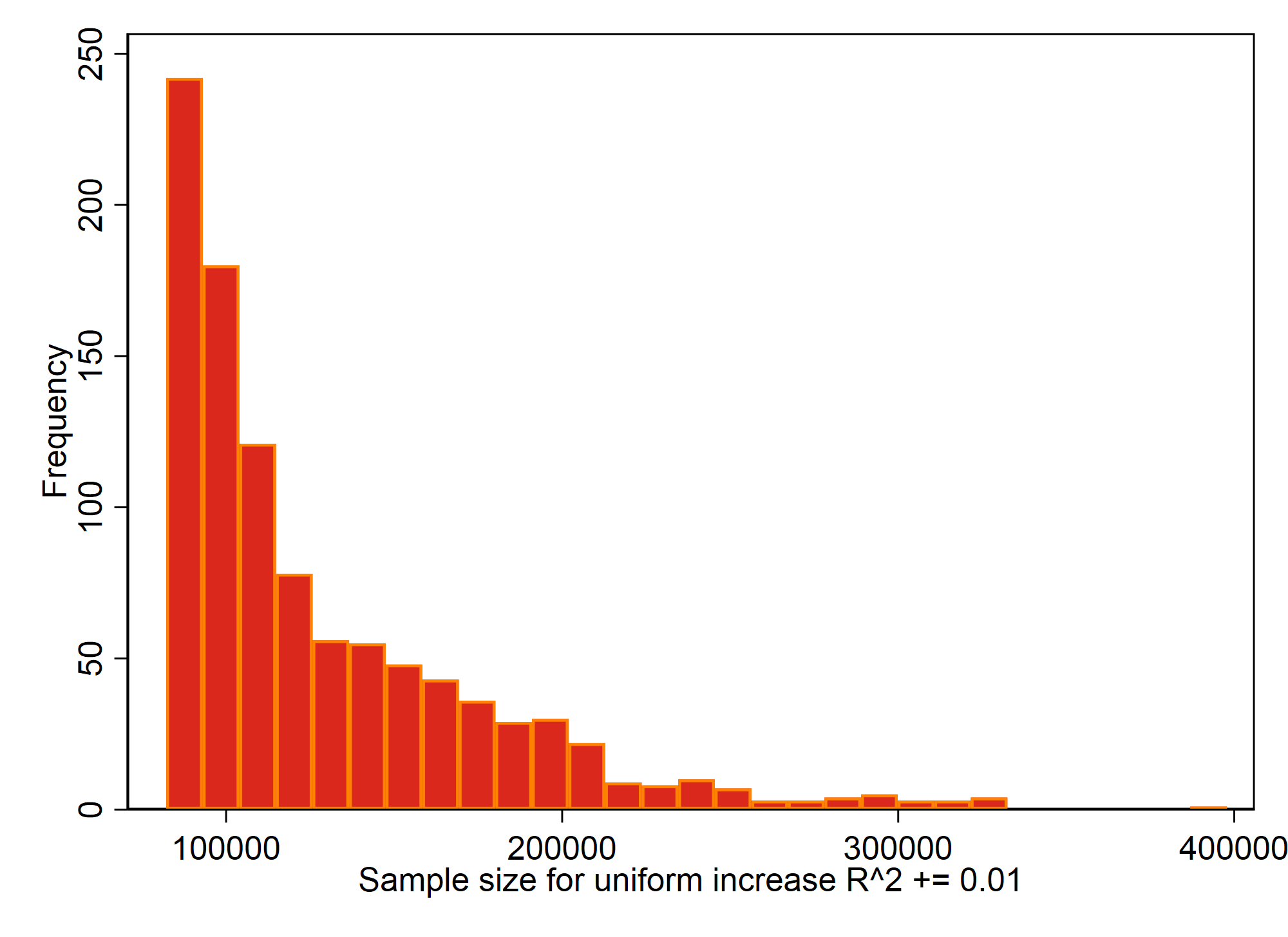}
        \\
        (a) & (b) \\
        \includegraphics[scale=0.09]{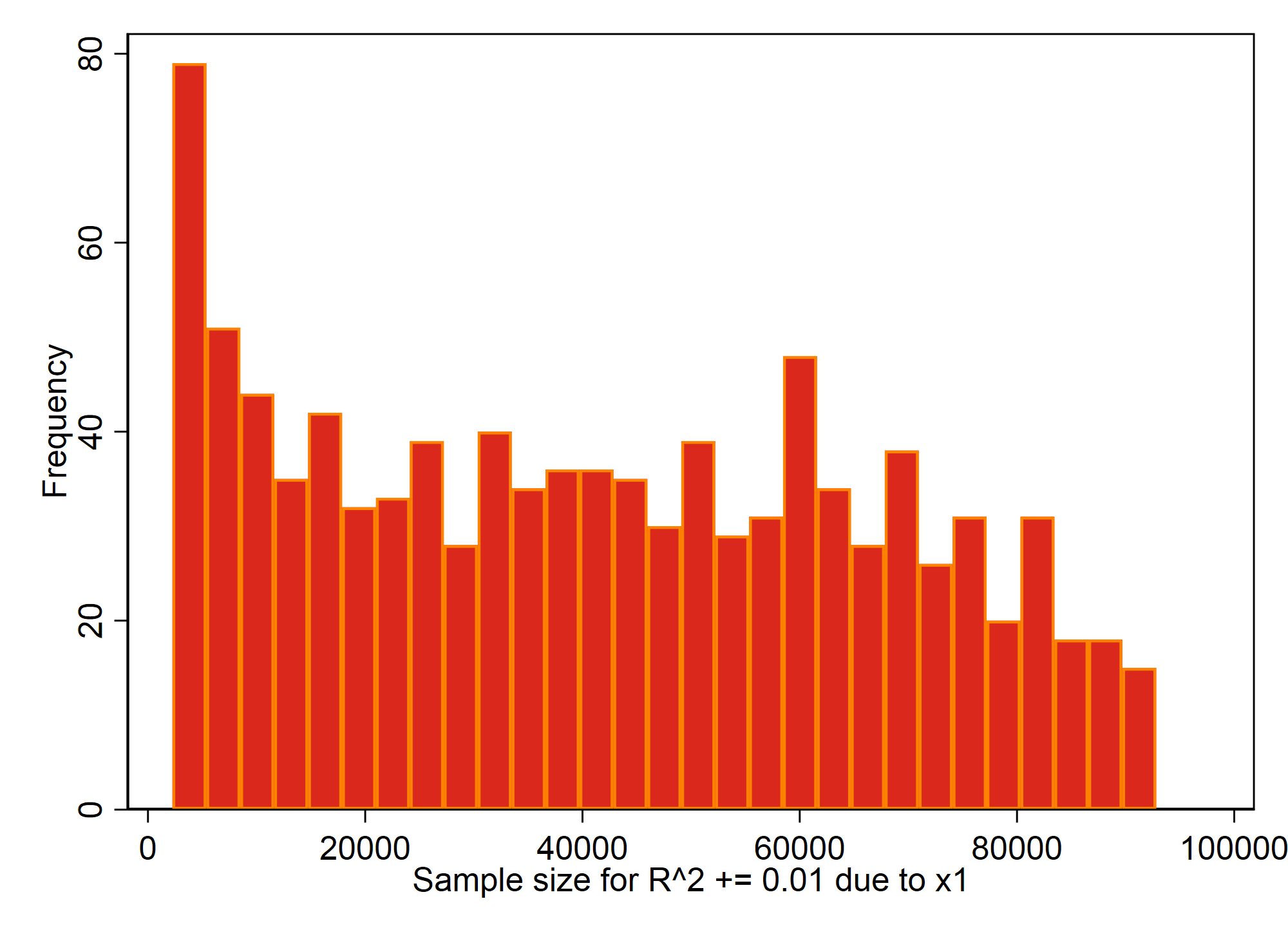}
        &
        \includegraphics[scale=0.09]{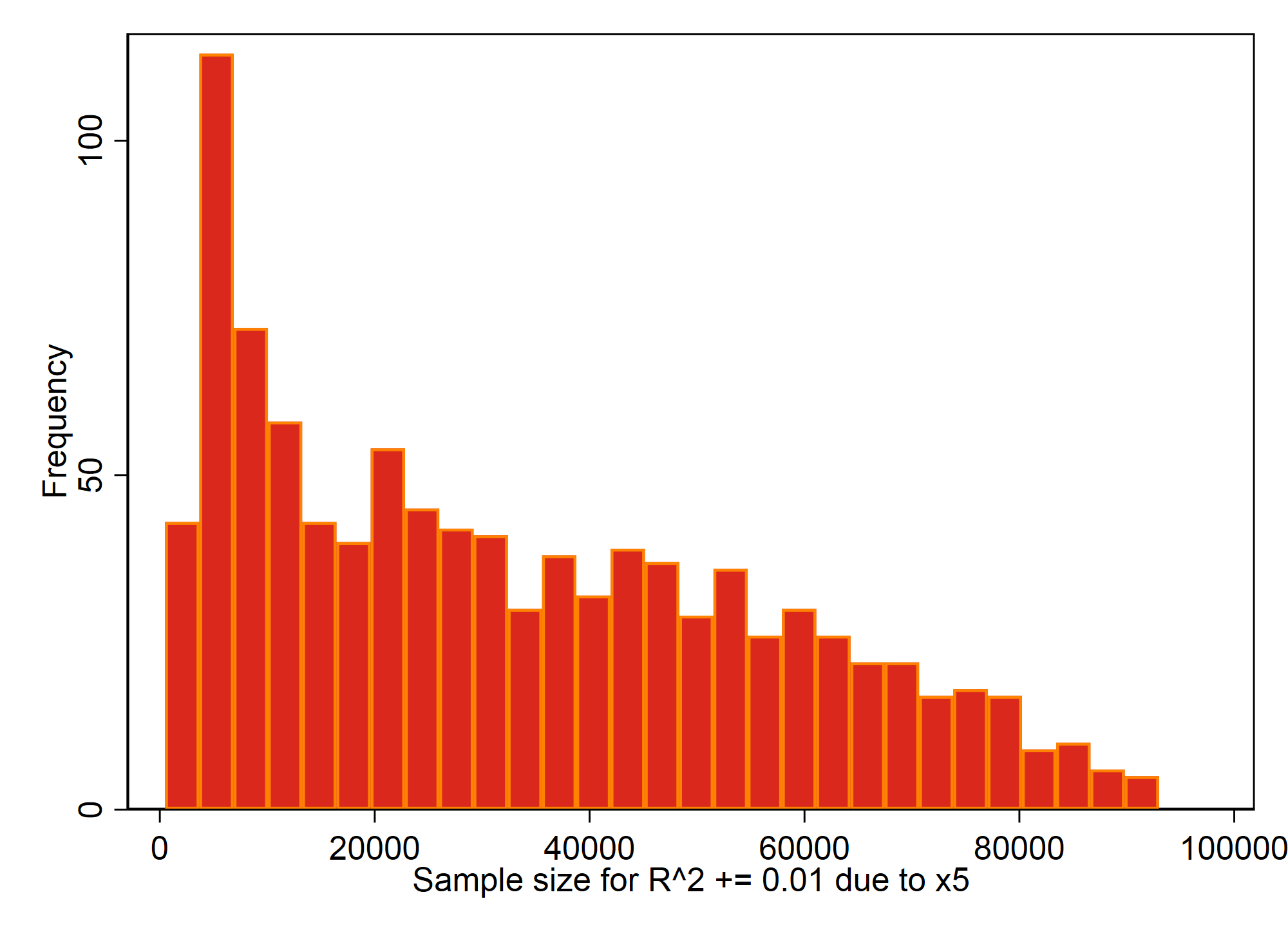}
        \\
        (c) & (d)
    \end{tabular}

    \caption{
        \label{fig:explore:n}
        Sample size to ensure the necessary detectable effect.
        (a) Overall test $H_0: R^2=0$;
        (b) $R^2$ increase due to overall explanatory power increase from $R^2=0.15$ by $0.01$;
        (c) $R^2$ increase due to an increase in explanatory power from $R^2=0.15$ by $0.01$ due to $x_1$;
        (d) $R^2$ increase due to an increase in explanatory power from $R^2=0.15$ by $0.01$ due to $x_5$.
    }

\end{figure}

1,000 Monte Carlo draws of the $\beta$ vector, and subsequent analytical computation of asymptotic variances and power, were done.
Results are presented graphically. Figure \ref{fig:explore:n} presents the sample sizes obtained in steps
\ref{enum:explor:n:overall}--\ref{enum:explor:n:poke} of the parameter exploration. A striking feature of the plot
is wide variability of the sample sizes as a function of the specific configuration of parameters. While the lower limit
of the sample size necessary to detect an overall increase in $R^2$ by $0.01$ is about $n=82K$, the median value
is $n=110K$, the 95th percentile is $n=220K$, and the maximum (worst case scenario) identified in this simulation is
$n=400K$. The patterns of the coefficients of the worst case scenarios typically indicate large coefficients of opposite signs
of the positively correlated variables ($x_1$ through $x_4$), or large coefficients of similar size of one of the positively
correlated factors ($x_1$ through $x_4$) and a high value of factor $x_5$ that is negatively correlated with all other subscales.
This wide range of variability makes it difficult to provide a definite recommendation concerning the sample
size for the study to the stakeholders. A conservative value based on a high percentile (80\% or 90\%) can be recommended,
to protect against bad population values of regression parameters at the expense of a potentially unnecessary increase in costs.

\begin{figure}[!bh]

    \centering

    \includegraphics[scale=0.09]{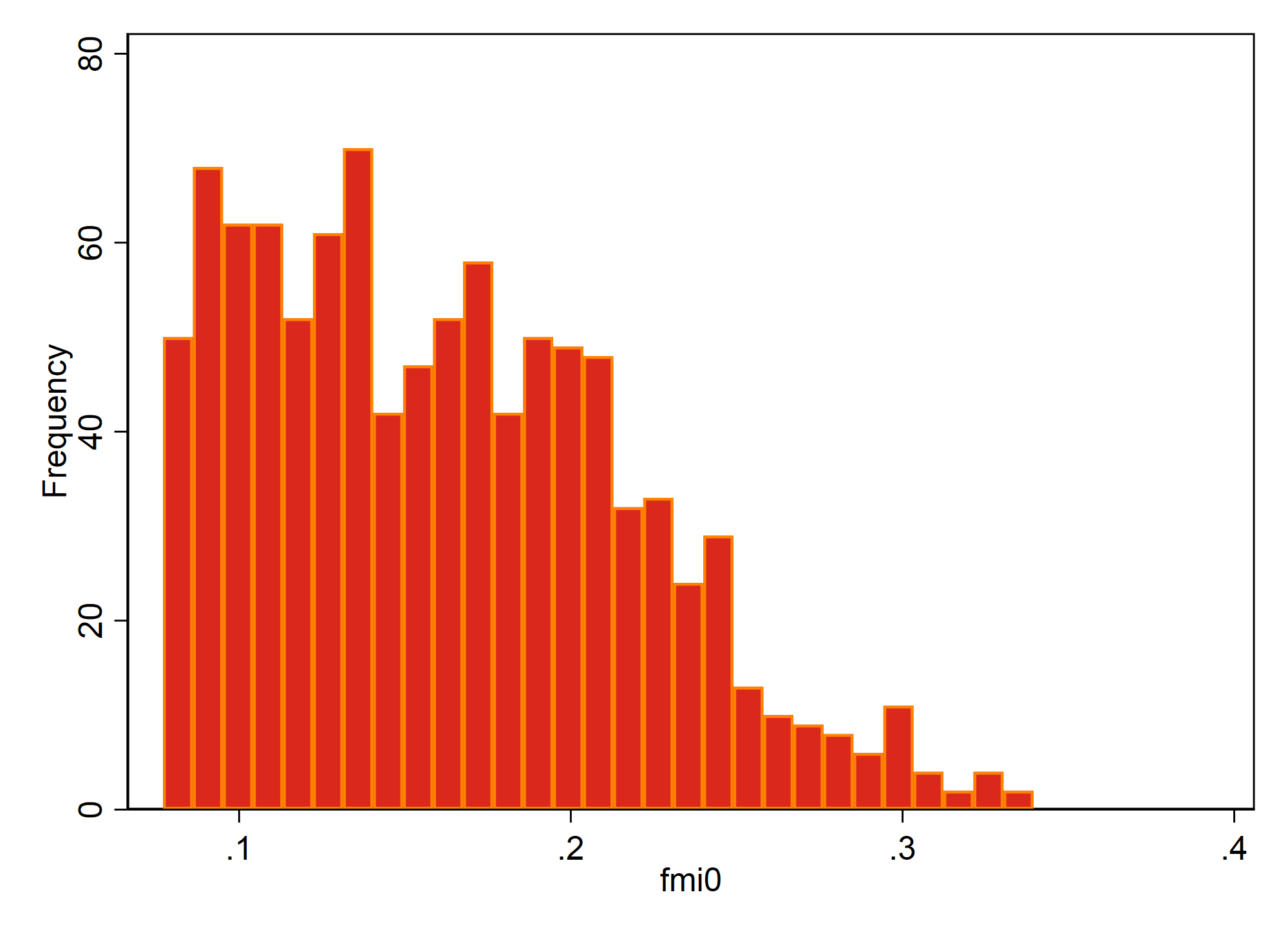}

    (a)

    \includegraphics[scale=0.09]{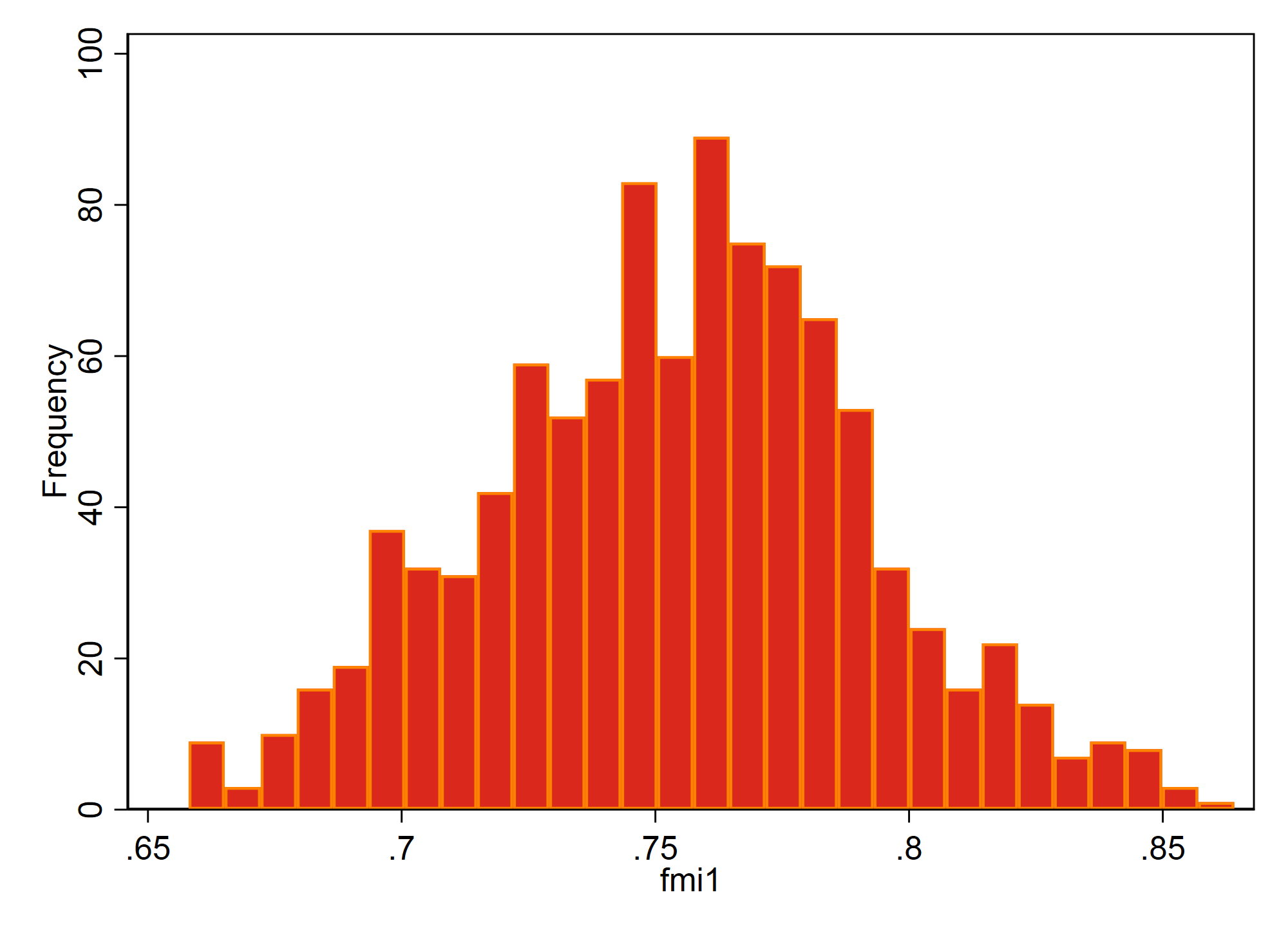}

    (b)

    \includegraphics[scale=0.09]{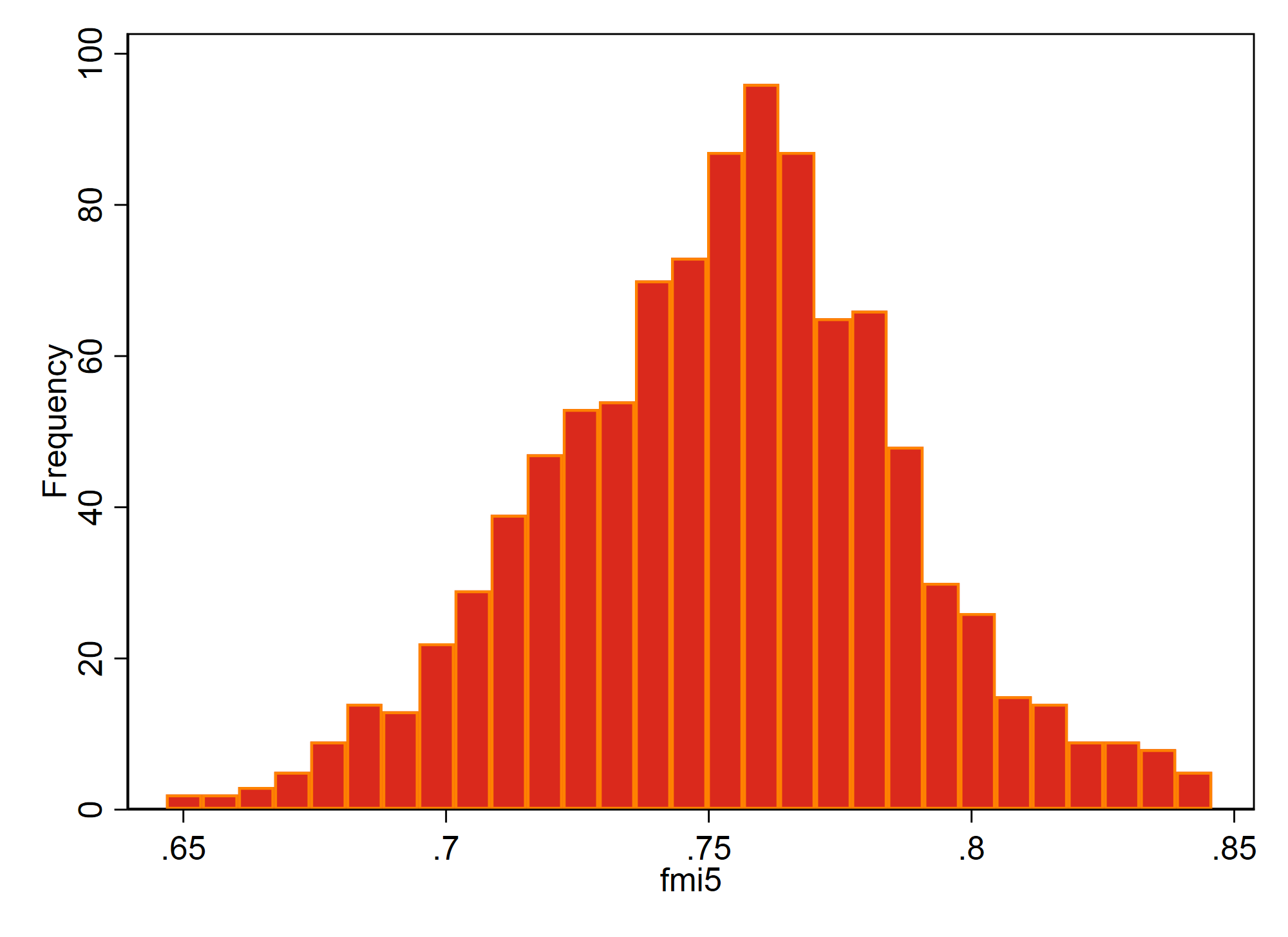}

    (c)

    \caption{
        \label{fig:explore:fmi}
        Fraction of missing information:
        (a) intercept; (b) slope of $x_1$; (c) slope of $x_5$.
    }

\end{figure}

Figure \ref{fig:explore:fmi} presents the exploration distribution of the fraction of missing information
due to the missing data. FMI for the intercept is generally low, below 0.2. FMI for regression slopes are generally high,
in the range of about 70\% to 80\%. Given the structure of the missing data shown by the
multiple matrix sampling design in Table \ref{tab:bigfive},
each of the predictor variables is observed in 40\% of the data (informing the diagonal entries of the $X'X$ matrix),
and each pairwise combination of the regressors is observed in 10\% of the data (informing the off-diagonal entries).
This yields an expected information loss for the predictor variables somewhere between 60\% and 90\%.

\section{Simulation 2: Performance in finite samples}

To study the performance of estimation methods based on SEM estimation with missing data, and on multiple imputation procedures,
a simulation with microdata was also performed. For each simulation draw, the following steps were taken.

\begin{enumerate}
    \item Sample size is set to $n=1,000$ (i.e., $100$ observations per form).
    \item Multivariate non-normal factor scores are simulated:
    \begin{enumerate}
        \item The non-normal principal components of $x_1, \ldots, x_5$ are simulated as
            \begin{align}
                f_1 & = -\ln u_1-1, \quad u_1 \sim U[0,1] \\
                f_2 & = (2 b-1) (-\ln u_2-1), \quad b \sim {\rm Bernoulli}(0.5), u_2 \sim U[0,1] \\
                f_3, f_4, f_4 & \sim N(0,1)
            \end{align}
            so that each principal component has a mean of 0 and variance of 1,
            with all the underlying random variables being drawn independently of each other.
            The first component $f_1$ has a marginal exponential distribution with a heavy right tail,
            ensuring the overall skewness of each factor. The second component has a bimodal distribution
            with two exponential components and heavy tails. The remaining three components are normal.
        \item The factor values are reconstructed as
            \begin{equation}
                \begin{pmatrix}
                    x_1 \\ x_2 \\ x_3 \\ x_4 \\ x_5
                \end{pmatrix}
                =
                \sum_{j=1}^5 \mathbf{u}_j \sqrt{\lambda_j} f_j
                \label{eq:Sigma:big5:eigenproblem}
            \end{equation}
            where $\Sigma_{\rm Big 5} = U' \Lambda U$ is the eigenvalue decomposition of the target
            covariance matrix (\ref{eq:bigfive:corr}) of the Big Five factors.
    \end{enumerate}
        \item The outcome is obtained as $y=0.3x_1 + 0.3x_4 + \varepsilon, \varepsilon \sim N(0,1.248602)$
            where the specific value of the residual variance was chosen to ensure that $R^2=0.15$ in the population.
        \item The regression model with the complete data is fit to obtain the benchmark for FMI calculation.
        \item The values of regressors were deleted in accordance with
            the multiple matrix sampling design in Table \ref{tab:bigfive}.
        \item The normal theory based SEM model for missing data was fit; regression parameter estimates
            and their asymptotic standard errors based on the inverse Hessian were recorded.
        \item $M=50$ complete data sets were imputed using multivariate normal imputation model.
        \item The regression model was estimated using the first $M=5$ data sets, in accordance with the traditional
            recommendation regarding the number of imputed data sets. Regression parameter estimates
            and their asymptotic standard errors based on the Rubin's rules were recorded.
        \item The regression model was estimated using all of the $M=50$ data sets. Regression parameter estimates
            and their asymptotic standard errors based on the Rubin's rules were recorded.
        \item $M=50$ complete data sets were imputed using predictive mean matching imputation model for each of the missing variables.
        \item The regression model was estimated using the first $M=5$ data sets, in accordance with the traditional
            recommendation regarding the number of imputed data sets. Regression parameter estimates
            and their asymptotic standard errors based on the Rubin's rules were recorded.
        \item The regression model was estimated using all of the $M=50$ data sets. Regression parameter estimates
            and their asymptotic standard errors based on the Rubin's rules were recorded.
\end{enumerate}

There were 1,200 Monte Carlo samples drawn.

\begin{figure}[!bh]
  \centering
  \includegraphics[scale=0.2]{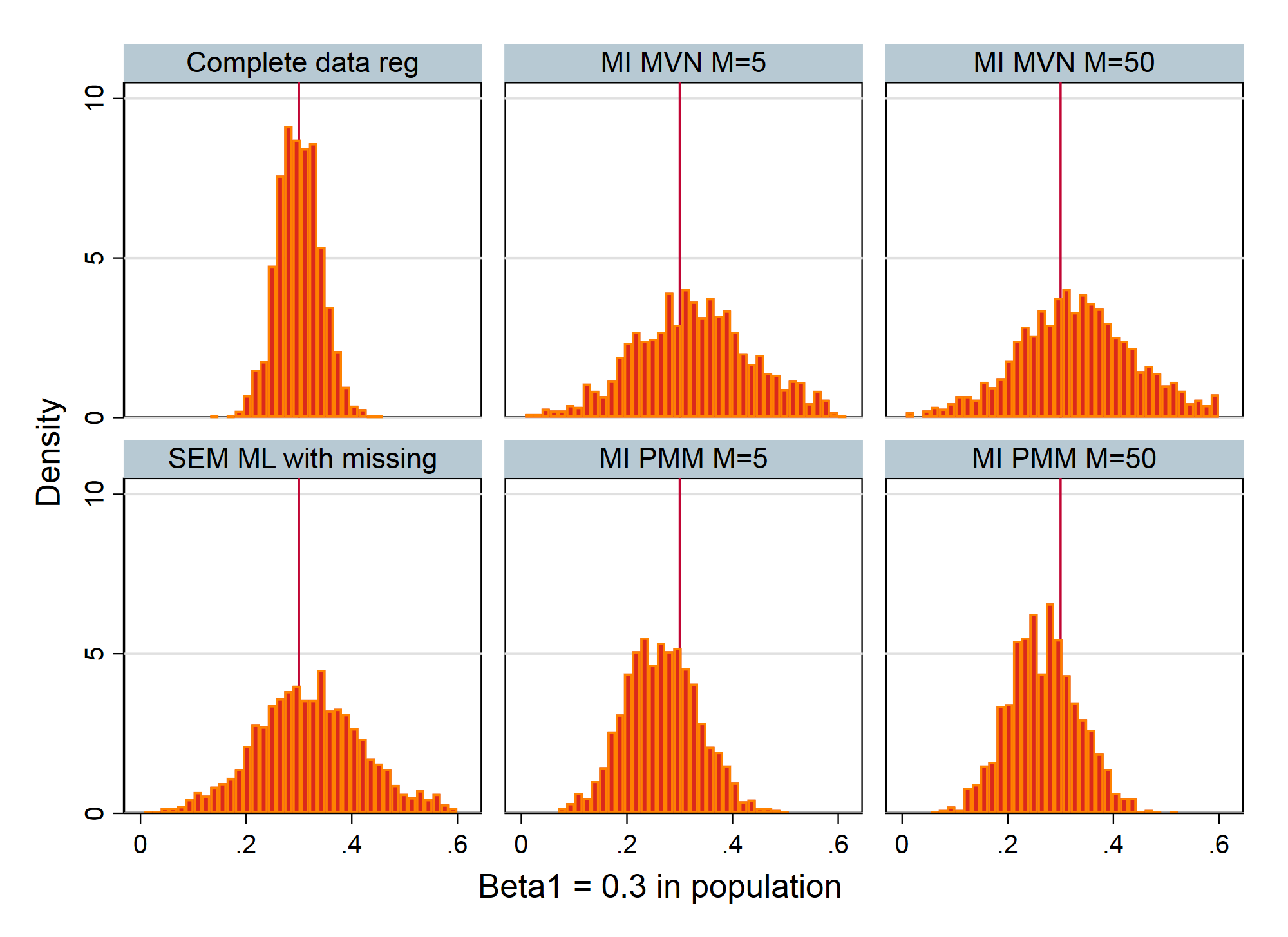}
  \caption{Sampling distributions of the parameter estimates $\hat\beta_1$ across different methods.}\label{fig:simul:b1}
\end{figure}

Figure \ref{fig:simul:b1} reports the simulated distributions of the estimates of parameter $\beta_1$.
The population value of 0.3 is shown as a vertical line on the plot. As expected, the complete data
regression model demonstrates higher efficiency. Estimates based on the multivarate normal methods
are biased up, while those based on MI with predictive mean matching are biased down.
Distributions of the estimates based on the multivariate normal methods are more spread out
than the asymptotic variance based on (\ref{eq:asvar:beta-hat}), while those based
on PMM MI are less spread out, with apparent efficiency gains extracted from higher moments of the data.
The plots in Figure \ref{fig:simul:b1} are truncated, with about 3\% of the Monte Carlo simulations
outside the right range of the plot (the value of $\beta_1=0.6$), and about 1\% of the Monte Carlo simulations
outside the left range of the plot (the value of $\beta_1=0$) for each of the methods based on multivariate normality.
Details for $\hat\beta_1$ and other regression coefficient estimates are provided in Table \ref{tab:simul:b1}.

% \begin{sidewaystable}[p]
\begin{table}[!th]
  \centering
  \caption{Monte Carlo means, [95\% confidence intervals] for the means and $\langle$standard deviations$\rangle$ for regression parameter estimates.}
  \label{tab:simul:b1}

  \begin{tabular}{l|ccccc}
     Method & $\hat\beta_1$ & $\hat\beta_2$ & $\hat\beta_3$ & $\hat\beta_4$ & $\hat\beta_5$ \\
     \hline 
     Complete & 0.3002	& 0.0016	& 0.0006	& 0.3002	& 0.0015  \\
     data & [0.298,0.303]	& [-0.001,0.004]	& [-0.002,0.003]	& [0.298,0.303]	& [-0.001,0.004] \\
     regression	& $\langle$ 0.0418	$\rangle$ & $\langle$ 0.0408 $\rangle$ & $\langle$ 0.0440 $\rangle$ & $\langle$ 0.0414 $\rangle$ & $\langle$ 0.0413 $\rangle$ 
     \medskip
     \\
     SEM with & 0.3277	& -0.0203	& -0.0130	& 0.3324	& 0.0096  \\
     MVN & [0.320,0.336]	& [-0.028,-0.013]	& [-0.022,-0.004]	& [0.324,0.340]	& [0.003,0.017]	\\
     missing data & $\langle$ 0.1414	$\rangle$ & $\langle$ 0.1356 $\rangle$ & $\langle$ 0.1588 $\rangle$ & $\langle$ 0.1429 $\rangle$ & $\langle$ 0.1249 $\rangle$ 
     \medskip
     \\
     MI using & 0.3369	& -0.0253	& -0.0173	& 0.3393	& 0.0091  \\
     MVN model, & [0.329,0.345]	& [-0.033,-0.017]	& [-0.027,-0.008]	& [0.331,0.347]	& [0.002,0.016] \\
     $M=5$ & $\langle$ 0.1390	$\rangle$ & $\langle$ 0.1415 $\rangle$ & $\langle$ 0.1645 $\rangle$ & $\langle$ 0.1435 $\rangle$ & $\langle$ 0.1259 $\rangle$ 
     \medskip
     \\
     MI using & 0.3430	& -0.0314	& -0.0208	& 0.3466	& 0.0109  \\
     MVN model, & [0.334,0.352]	& [-0.040,-0.023]	& [-0.031,-0.011]	& [0.338,0.355]	& [0.003,0.018] \\
     $M=50$ & $\langle$ 0.1556	$\rangle$ & $\langle$ 0.1507 $\rangle$ & $\langle$ 0.1760 $\rangle$ & $\langle$ 0.1531 $\rangle$ & $\langle$ 0.1336 $\rangle$
     \medskip
     \\
     MI using & 0.2661	& 0.0356	& 0.0261	& 0.2666	& -0.0056 \\
     PMM model, & [0.262,0.270]	& [0.032,0.039]	& [0.022,0.030]	& [0.263,0.271]	& [-0.009,-0.002] \\
     $M=5$ & $\langle$ 0.0707	$\rangle$ & $\langle$ 0.0679 $\rangle$ & $\langle$ 0.0758 $\rangle$ & $\langle$ 0.0707 $\rangle$ & $\langle$ 0.0631 $\rangle$ 
     \medskip
     \\
     MI using & 0.2678	& 0.0361	& 0.0251	& 0.2671	& -0.0043 \\
     PMM model, & [0.264,0.272]	& [0.032,0.040]	& [0.021,0.029]	& [0.263,0.271]	& [-0.008,-0.001] \\
     $M=50$ & $\langle$ 0.0676	$\rangle$ & $\langle$ 0.0656 $\rangle$ & $\langle$ 0.0719 $\rangle$ & $\langle$ 0.0665 $\rangle$ & $\langle$ 0.0591 $\rangle$ \\
     \hline
     Population & 0.3 & 0 & 0 & 0.3 & 0 \\
     & $\langle$ 0.0791	$\rangle$ & $\langle$ 0.0856 $\rangle$ & $\langle$ 0.0926 $\rangle$ & $\langle$ 0.0824 $\rangle$ & $\langle$ 0.0832 $\rangle$ \\
  \end{tabular}
% \end{sidewaystable}					
\end{table}

\begin{figure}[!bh]
  \centering
  \includegraphics[scale=0.2]{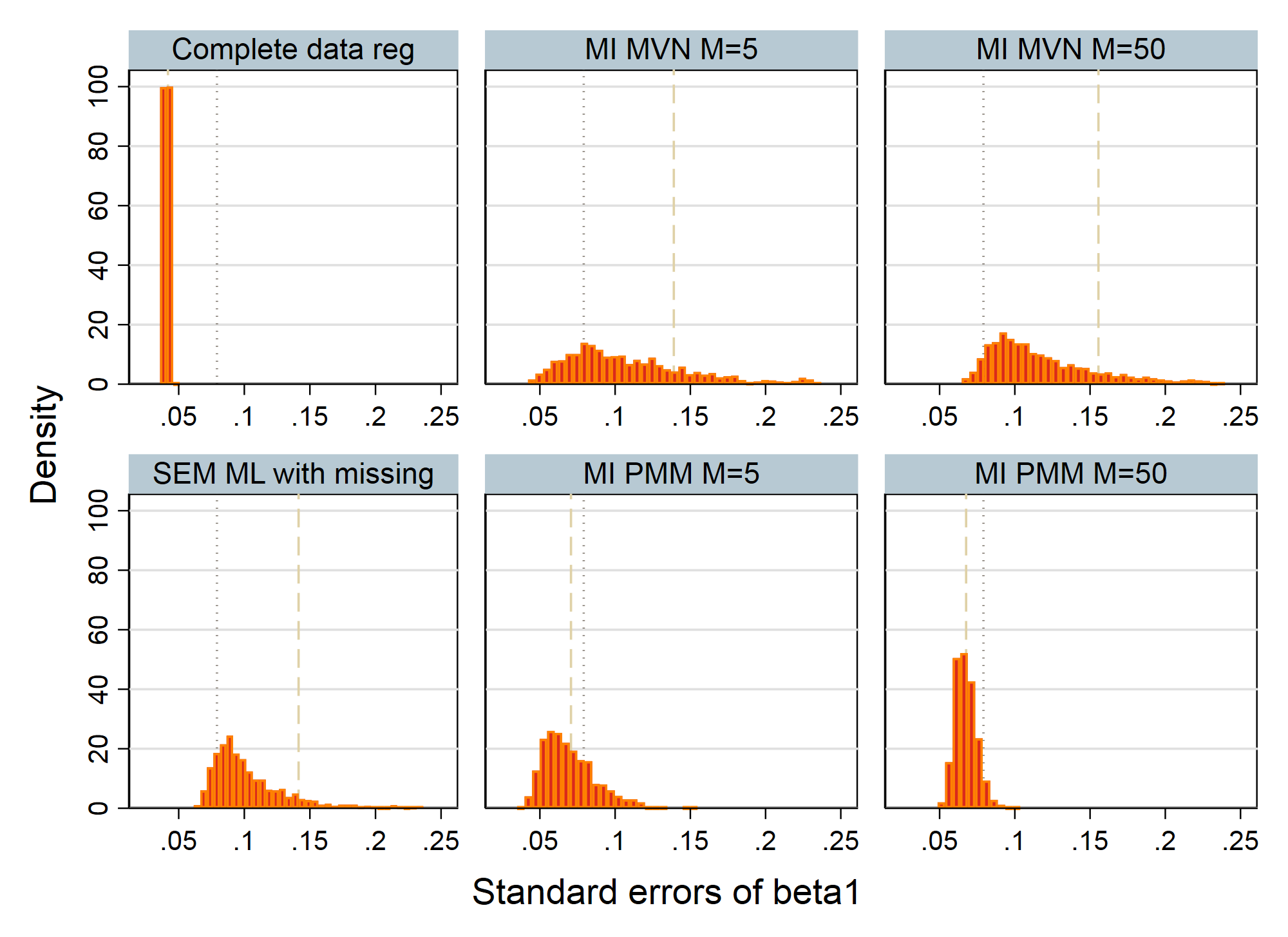}
  \caption{Sampling distributions of the standard errors of $\hat\beta_1$ across different methods.}\label{fig:simul:se1}
\end{figure}

Figure \ref{fig:simul:se1} provides the Monte Carlo distributions of the standard errors reported for the missing data methods.
The dotted vertical line is the asymptotic standard error based on (\ref{eq:asvar:beta-hat}), 0.0791. The dashed lines are empirical
means of the standard errors. All distributions are skewed with heavy right tails. The distributions of the standard errors
based on multivariate data contain outliers outside the range of the plot (3\% of the SEM with missing data results; 6\% of
the results for MI using the multivariate normal model with $M=5$; 8\% of the results for MI using the multivariate normal model with $M=50$;
the range of the plots is from 0 to $3\times$ the asymptotic standard error, 0.0791). Distributions of the standard errors
for the multivariate normal methods are significantly higher that this asymptotic standard error, which reflects, to some extent,
the greater variability of the estimates observed above in Figure \ref{fig:simul:b1} and Table \ref{tab:simul:b1}.
Distributions of the standard errors
for the PMM MI method are significantly lower that the asymptotic standard error, which reflects, to some extent,
the lower variability of the estimates based on this method.
A higher number of multiple imputations $M=50$ vs. $M=5$ helps to stabilize the variance estimates, particularly in the case of PMM.

Coverage of the nominal 95\% confidence intervals is analyzed in Table \ref{tab:simul:cover95}. Despite the shortcomings
of both the point estimates and the standard errors noted above, things seem to balance out and provide confidence interval coverage
fairly close to the target.

\begin{table}[!th]
  \centering
  \caption{Coverage of the nominal 95\% coverage intervals.}\label{tab:simul:cover95}

  \begin{tabular}{l|ccccc}
     Method & $\hat\beta_1$ & $\hat\beta_2$ & $\hat\beta_3$ & $\hat\beta_4$ & $\hat\beta_5$ \\
     \hline \\
     Complete data regression	 & 95.5\% & 95.4\% & 95.1\% & 95.8\% & 97.3\% \\
     SEM with MVN missing data	& 97.8\% & 98.6\% & 97.3\% & 98.7\% & 98.4\% \\
     MI using MVN model, $M=5$	 &93.3\% & 93.8\% & 93.0\% & 93.3\% & 96.8\% \\
     MI using MVN model, $M=50$	&92.8\% & 93.4\% & 93.2\% & 93.1\% & 97.9\% \\
     MI using PMM model, $M=5$	 &94.4\% & 95.6\% & 94.5\% & 95.0\% & 94.2\% \\
     MI using PMM model, $M=50$	&96.3\% & 96.6\% & 95.8\% & 96.8\% & 97.2\% \\
  \end{tabular}
\end{table}

\begin{figure}[!bh]
  \centering
  \includegraphics[scale=0.2]{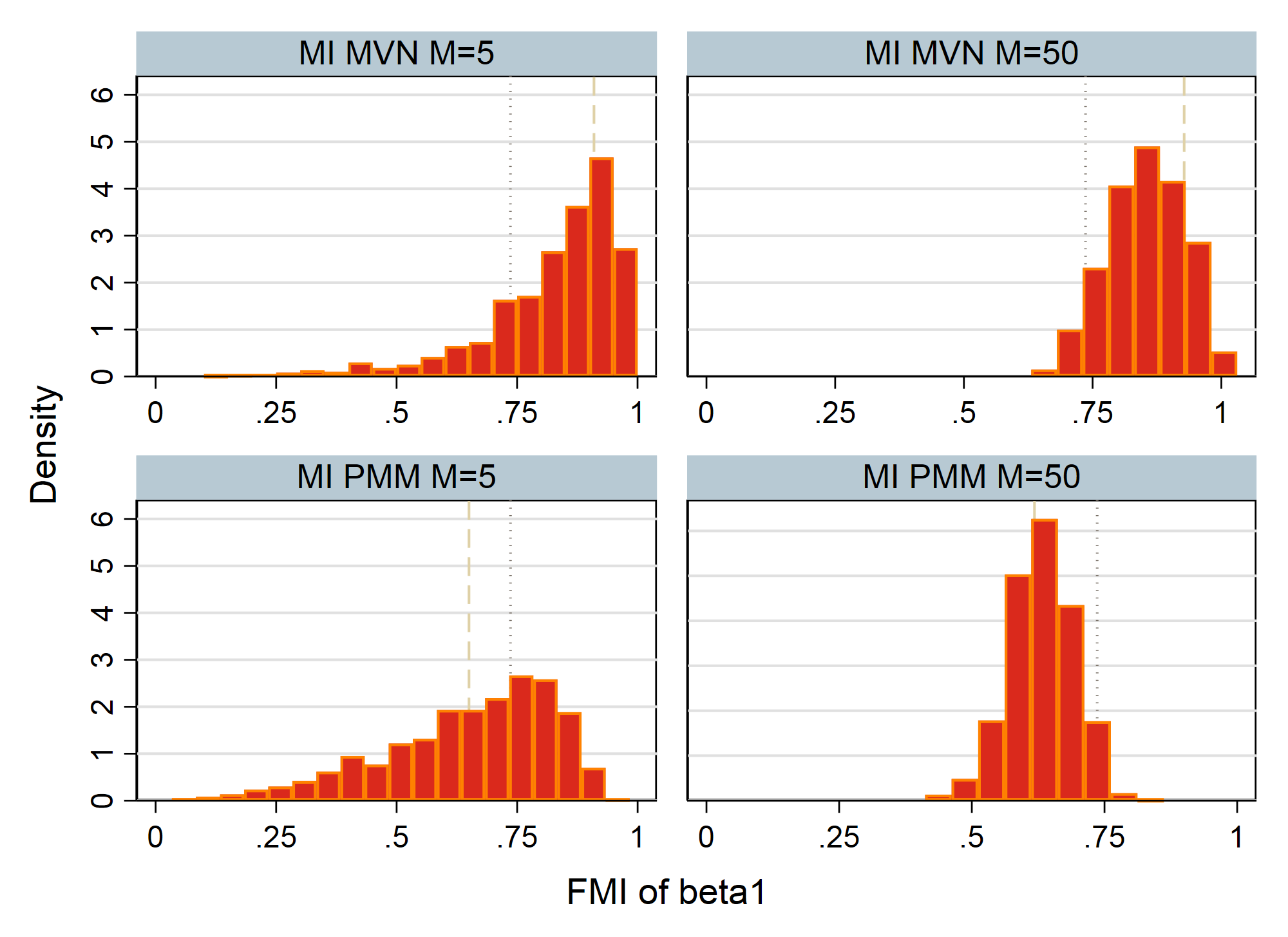}
  \caption{Reported fraction of missing information.}\label{fig:simul:fmi}
\end{figure}

Estimated fractions of missing information reported by the software are shown on Figure \ref{fig:simul:fmi}.
The dotted line is the value based on asymptotic variance, 73.6\%. Dashed lines are the empirical FMI,
based on the ratios of the Monte Carlo variance of $\hat\beta_1$ based on a given missing data method
to the variance of $\hat\beta_1$ based on the complete data. The latter empirical FMI is greater than
the theoretical one for the MI methods based on the multivariate normality assumption, and lower than
the theoretical one for the PMM MI methods. The methods based on multivariate normality appear to underestimate
FMI, as the distributions of the reported empirical FMI appear to the left of the true value (dashed line).
The FMI that come out of PMM MI appear to be more accurate. An increase in the number of completed data sets
from $M=5$ to $M=50$ helps to improve stability of the FMI estimates, making the distributions of the empirical FMI
more concentrated.

\section{Concluding remarks}

This paper provides an analytical framework for analysis of regression models
(and, more generally, other statistical methods that are based on the covariance matrices of observed
items or scales)
that allows for quick power analysis avoiding computationally intensive simulations.

Revisiting the initial motivation of burden reduction, the results are underwhelming.
Is burden really reduced by multiple matrix sampling in the example considered?
Out of five explanatory variables (based on approximately 8 survey items each) and one outcome,
only three variables are collected on each of the matrix sampled instrument forms.
This translates to about 50\% burden reduction per respondent. However,
given that the loss of information quantified by the fraction of missing information
(FMI) is about 75–-80\%, the data collection sample sizes would
need to be about 4--5 times larger compared to the traditional data collection
of all items at once.
Unless the response rate drops sharply by a factor of more than two due to
the increase in questionnaire length, the total public burden is increased.

The sample sizes necessary to detect the required effect sizes in increased $R^2$
demonstrate long tails in the exploration of parameter spaces. These long tails
make it difficult to plan for the worst-case scenarios associated with ``unfortunate''
regression parameter configurations. Should a specific decision need to be made
based on the parameter explorations akin to those undertaken in Section \ref{sec:explore},
the trade-off between the survey costs due to large sample sizes and risks of having
an underpowered study should the coefficient estimates be found to have an ``unfortunate''
configuration should be carefully discussed with the survey stakeholders to find
the most appropriate course of action.

We conducted a finite sample simulation with non-normal data and several missing data methods,
and determined that the methods that assume multivariate normality generally perform poorly,
and generate a non-negligible proportion of really bad outliers. In comparison,
semiparametric multiple imputation by predictive mean matching with sufficiently large
number of imputed data sets seem to work best.

Our work can be extended in a number of additional dimensions. The derivations of asymptotic
variances are based on the working assumption of multivariate normality and using
the inverse information matrix to estimate variances. With non-normal data, the problem
can be formulated in terms of estimating equations, and sandwich variance estimators
should be formed. As our simulation demonstrated, asymptotic standard errors based on
inverse information matrix are inadequate for the analysis methods that we used,
leading to underestimates with misspecified normality-based methods, and overestimates with
a more accurate semiparametric method.

The current paper assumed independence of respondents. In practice, complex survey features
such as strata, clusters, unequal probabilities of selection, and weight calibration would
affect asymptotic properties of the estimates. In particular, the sandwich variance estimation
will be required. Many practical survey statistics issues may also interact with multiple
matrix sampling in unusual ways. How would differential nonresponse by form affect the results?
What should we do when a stratum has fewer than two cases of a given form? These and other
questions related to design-based inference would need to be answered when multiple matrix sampling
is applied in practice.

Finally, in terms of ensuring adequate measurement properties, we note
that psychometric properties are usually established and validated for scales,
but not necessarily subscales that respondents are exposed to in multiple matrix sampling instruments.
In particular, if the order of the items, or the degree of mixing of items from the different
subscales of the Big Five Inventory is important for the validity of the scale and its subscales,
these properties may be violated when shorter subscales are administered that require
the respondent to answer similar questions more frequently.

\clearpage

\bibliography{everything}
\bibliographystyle{agsm}

\end{document}